\documentclass[11pt,a4paper,DIV12]{scrartcl}

\usepackage{MCMaxTSP}

\title{Approximating Multi-Criteria Max-TSP\thanks{An extended abstract of this
       work will appear in \emph{Proc.\ of the 16th Ann.\ European Symposium on
       Algorithms (ESA 2008)}.}}

\author{Markus Bl\"aser \and Bodo Manthey \and Oliver Putz}

\date{\small Saarland University, Computer Science \\
Postfach 151150, 66041 Saarbr\"ucken, Germany \\
\texttt{blaeser/manthey@cs.uni-sb.de}, \texttt{oli.putz@gmx.de}}

\begin{document}


\maketitle


\begin{abstract}
We present randomized approximation algorithms for multi-criteria \maxtsp. For
\maxstsp\ with $k > 1$ objective functions, we obtain an approximation ratio of
$\frac 1k - \eps$ for arbitrarily small $\eps > 0$. For \maxatsp\ with $k$
objective functions, we obtain an approximation ratio of $\frac{1}{k+1} - \eps$.
\end{abstract}


\section{Multi-Criteria Traveling Salesman Problem}
\label{sec:mctsp}

\subsection{Traveling Salesman Problem}
\label{ssec:atsp}

The traveling salesman problem (TSP) is one of the most fundamental problems in
combinatorial optimization. Given a graph, the goal is to find a Hamiltonian
cycle of minimum or maximum weight. We consider finding Hamiltonian cycles of
maximum weight (\maxtsp).

An instance of \maxtsp\ is a complete graph $G=(V,E)$ with edge weights
$w: E \to \nat$. The goal is to find a Hamiltonian cycle of maximum weight. The
weight of a Hamiltonian cycle (or, more general, of a subset of $E$) is the sum
of the weights of its edges. If $G$ is undirected, we speak of \maxstsp\
(symmetric TSP). If $G$ is directed, we have \maxatsp\ (asymmetric TSP).

Both \maxstsp\ and \maxatsp\ are NP-hard and APX-hard. Thus, we are in need of
approximation algorithms. The currently best approximation algorithms for
\maxstsp\ and \maxatsp\ achieve approximation ratios of $61/81$ and $2/3$,
respectively~\cite{ChenEA:ImprovedMaxTSP:2005,KaplanEA:TSP:2005}.

Cycle covers are an important tool for designing approximation algorithms for
the TSP. A cycle cover of a graph is a set of vertex-disjoint cycles such that
every vertex is part of exactly one cycle. Hamiltonian cycles are special cases
of cycle covers that consist of just one cycle. Thus, the weight of a
maximum-weight cycle cover is an upper bound for the weight of a maximum-weight
Hamiltonian cycle. In contrast to Hamiltonian cycles, cycle covers of minimum or
maximum weight can be computed efficiently using matching
algorithms~\cite{AhujaEA:NetworkFlows:1993}.

\subsection{Multi-Criteria Optimization}
\label{ssec:mcopt}

In many optimization problems, there is more than one objective function.
Consider buying a car: We might want to buy a cheap, fast car with a good gas
mileage. How do we decide which car suits us best? With multiple criteria
involved, there is no natural notion of a best choice. Instead, we have to be
content with a trade-off. The aim of multi-criteria optimization is to cope with
this problem. To transfer the concept of an optimal solution to multi-criteria
optimization problems, the notion of \emph{Pareto curves} was introduced
(cf.\ Ehrgott~\cite{Ehrgott:MulticriteriaOpt:2005}). A Pareto curve is a
set of solutions that can be considered optimal.

More formally, a $k$-criteria optimization problem consists of instances $I$,
solutions $\sol(X)$ for every instance $X \in I$, and $k$ objective functions
$w_1, \ldots, w_k$ that map $X \in I$ and $Y \in \sol(X)$ to $\nat$. Throughout
this paper, our aim is to maximize the objective functions. We say that a
solution $Y \in \sol(X)$ \emph{dominates} another solution $Z \in \sol(X)$ if
$w_i(Y, X) \geq w_i(Z, X)$ for all $i \in [k] = \{1, \ldots, k\}$ and
$w_i(Y, X) > w_i(Z, X)$ for at least one $i$. This means that $Y$ is strictly
preferable to $Z$. A \emph{Pareto curve} (also known as \emph{Pareto set} or
\emph{efficient set}) for an instance contains all solutions of that instance
that are not dominated by another solution.

Unfortunately, Pareto curves cannot be computed efficiently in many cases:
First, they are often of exponential size. Second, because of straightforward
reductions from knapsack problems, they are NP-hard to compute even for
otherwise easy problems. Thus, we have to be content with approximate Pareto
curves.

For simpler notation, let $w(Y,X) = (w_1(Y,X), \ldots, w_k(Y,X))$. We will omit
the instance $X$ if it is clear from the context. Inequalities are meant
component-wise. A set $\mathcal P \subseteq \sol(X)$ of solutions is called an
\emph{$\alpha$ approximate Pareto curve} for $X \in I$ if the following holds:
For every solution $Z \in \sol(X)$, there exists a $Y \in \mathcal P$ with
$w(Y) \geq \alpha w(Z)$. We have $\alpha \leq 1$, and a $1$ approximate Pareto
curve is a Pareto curve. (This is not precisely true if there are several
solutions whose objective values agree. However, in our case this is
inconsequential, and we will not elaborate on this for the sake of clarity.) An
algorithm is called an \emph{$\alpha$ approximation algorithm} if, given the
instance $X$, it computes an $\alpha$ approximate Pareto curve. It is called a
randomized $\alpha$ approximation algorithm if its success probability is at
least $1/2$. This success probability can be amplified to $1-2^{-m}$ by
executing the algorithm $m$ times and taking the union of all sets of solutions.
(We can also remove solutions from this union that are dominated by other
solutions in the union, but this is not required by the definition of an
approximate Pareto curve.)

Papadimitriou and Yannakakis~\cite{PapadimitriouYannakakis:TradeOffs:2000}
showed that $(1-\eps)$ approximate Pareto curves of size polynomial in the
instance size and $1/\eps$ exist. The technical requirement for the existence is
that the objective values of solutions in $\sol(X)$ are bounded from above by
$2^{p(N)}$ for some polynomial $p$, where $N$ is the size of $X$. This is
fulfilled in most natural optimization problems and in particular in our case.

A \emph{fully polynomial time approximation scheme} (FPTAS) for a multi-criteria
optimization problem computes $(1-\eps)$ approximate Pareto curves in time
polynomial in the size of the instance and $1/\eps$ for all $\eps > 0$.
Papadimitriou and Yannakakis~\cite{PapadimitriouYannakakis:TradeOffs:2000},
based on a result of Mulmuley et al.~\cite{MulmuleyEA:Matching:1987}, showed
that multi-criteria minimum-weight matching admits a \emph{randomized FPTAS},
i.~e., the algorithm succeeds in computing a $(1-\eps)$ approximate Pareto curve
with constant probability. This randomized FPTAS yields also a randomized FPTAS
for the multi-criteria maximum-weight cycle cover
problem~\cite{MantheyRam:MultiCritTSP:XXXX}, which we will use in the
following.

Manthey and Ram~\cite{MantheyRam:MultiCritTSP:XXXX,Manthey:MCATSP:2007} designed
randomized approximation algorithms for several variants of multi-criteria
Min-TSP. However, they leave it as an open problem to design any approximation
algorithm for \maxtsp.

\subsection{New Results}
\label{ssec:new}

We devise the first approximation algorithm for multi-criteria Max-TSP. For
$k$-criteria \maxstsp, we achieve an approximation ratio of $\frac 1k - \eps$
for arbitrarily small $\eps > 0$. For $k$-criteria \maxatsp, we achieve
$\frac{1}{k+1} - \eps$. Our algorithm is randomized. Its running-time is
polynomial in the input size and $1/\eps$ and exponential in the number $k$ of
criteria. However, the number of different objective functions is usually a
small constant.

The main ingredient for our algorithm is a decomposition technique for cycle
covers and a reduction from $k$-criteria instances to $(k-1)$-criteria
instances.

\section{Outline and Idea}
\label{sec:idea}

A straight-forward $1/2$ approximation for mono-criterion Max-ATSP is the
following: First, we compute a maximum-weight cycle cover $C$. Then we remove the
lightest edge of each cycle, thus losing at most half of $C$'s weight. In this
way, we obtain a collection of paths. Finally, we add edges to connect the paths
to get a Hamiltonian cycle. For Max-STSP, the same approach yields a $2/3$
approximation since the length of every cycle is at least three.

Unfortunately, this does not generalize to multi-criteria \maxtsp\ for which
``lightest edge'' is usually not well defined: If we break an edge that has
little weight with respect to one objective, we might lose a lot of weight with
respect to another objective. Based on this observation, the basic idea behind
our algorithm and its analysis is the following case distinction:

\begin{description}
\item[Light-weight edges:] If all edges of our cycle cover contribute only
      little to its weight, then removing one edge does not decrease the overall
      weight by too much. Now we choose the edges to be removed such that no
      objective loses too much of its weight.

\item[Heavy-weight edges:] If there is one edge that is very heavy with respect
      to at least one objective, then we take only this edge from the cycle
      cover. In this way, we have enough weight for one objective, and we
      proceed recursively on the remaining graph with $k-1$ objectives.
\end{description}

In this way, the approximation ratio for $k$-criteria \maxtsp\ depends on two
questions: First, how well can we decompose a cycle cover consisting solely of
light-weight edges? Second, how well can $(k-1)$-criteria \maxtsp\ be
approximated? We deal with the first question in Section~\ref{sec:decomp}. In
Section~\ref{sec:alg}, we present and analyze our approximation algorithms,
which also gives an answer to the second question. Finally, we give evidence
that the analysis of the approximation ratios is tight and point out some
ideas that might lead to better approximation ratios
(Section~\ref{sec:remarks}).

\section{Decompositions}
\label{sec:decomp}

Let $\alpha \in (0,1]$, and let $C$ be a cycle cover. We call a collection
$P \subseteq C$ of paths an \emph{$\alpha$-decomposition of $C$} if
$w(P) \geq \alpha w(C)$. (Remember that all inequalities are meant
component-wise.) In the following, our aim is to find
$\alpha$-decompositions of cycle covers consisting solely of light-weight
edges, that is, $w(e) \leq \alpha w(C)$ for all $e \in C$.

Of course, not every cycle cover possesses an $\alpha$-decomposition for every
$\alpha$. For instance, a single directed cycle of length two, where each edge
has a weight of $1$ shows that $\alpha = 1/2$ is best possible for a single
objective function in directed graphs. On the other hand, by removing the
lightest edge of every cycle, we obtain a $1/2$-decomposition.

For undirected graphs and $k = 1$, $\alpha = 2/3$ is optimal: We can find
a $2/3$-decomposition by removing the lightest edge of every cycle, and a single
cycle of length three, where each edge weight is $1$, shows that this is tight.

More general, we define $\alpha_k^d \in (0,1]$ to be the maximum number such
that every directed cycle cover $C$ with $w(e) \leq \alpha_k^d \cdot w(C)$ for
all $e \in C$ possesses an $\alpha_k^d$-decomposition. Analogously,
$\alpha_k^u \in (0,1]$ is the maximum number such that every undirected cycle
cover $C$ with $w(e) \leq \alpha_k^u \cdot w(C)$ possesses an
$\alpha_k^u$-decomposition. We have $\alpha_1^d = \frac 12$ and
$\alpha_1^u = \frac 23$, as we have already argued above. We also have
$\alpha_k^u \geq \alpha_k^d$ and $\alpha_k^u \leq \alpha_{k-1}^u$ as well as
$\alpha_k^d \leq \alpha_{k-1}^d$.

\subsection{Existence of Decompositions}
\label{ssec:existdecomp}

In this section, we investigate for which values of $\alpha$ such
$\alpha$-decompositions exist. In the subsequent section, we show how to
actually find good decompositions.  We have already dealt with $\alpha_1^u$ and
$\alpha_1^d$. Thus, $k \geq 2$ remains to be considered in the following
theorems. In particular, only $k \geq 2$ is needed for the analysis of our
algorithms.

Let us first normalize our cycle covers to make the proofs in the following a
bit easier. For directed cycle covers $C$, we can restrict ourselves to cycles
of length two: If we have a cycle $c$ of length $\ell$ with edges
$e_1, \ldots, e_\ell$, we replace it by $\lfloor \ell/2 \rfloor$ cycles
$(e_{2j-1}, e_{2j})$ for $j = 1, \ldots, \lfloor \ell/2 \rfloor$. If $\ell$ is
odd, then we add a edge $e_{\ell+1}$ with $w(e_{\ell+1}) = 0$ and add the cycle
$(e_\ell, e_{\ell+1})$. (Strictly speaking, edges are 2-tuples of vertices, and
we cannot simply reconnect them. What we mean is that we remove the edges of the
cycle and create new edges with the same names and weights together with
appropriate new vertices.) We do this for all cycles of length at least three
and call the resulting cycle cover $C'$. Now any $\alpha$-decomposition $P'$ of
the new cycle cover $C'$ yields an $\alpha$-decomposition $P$ of the original
cycle cover $C$ by removing the newly added edges $e_{\ell+1}$: In $C$, we have
to remove at least one edge of the cycle $c$ to obtain a decomposition. In $C'$,
we have to remove at least $\lfloor \ell/2 \rfloor$ edges of $c$, thus at least
one. Furthermore, if $w(e) \leq \alpha \cdot w(C)$ for every $e \in C$, then
also $w(e) \leq \alpha \cdot w(C')$ for every $e \in C'$ since we kept all edge
weights. This also shows $w(P) = w(P')$.

We are interested in $\alpha$-decompositions that work for all cycle covers with
$k$ objective functions. Thus in particular, we have to be able to decompose
$C'$. The consequence is that if every directed cycle cover that consists solely
of cycles of length two possesses an $\alpha$-decomposition, then every directed
cycle cover does so.

For undirected cycle covers, we can restrict ourselves to cycles of length
three: We replace a cycle $c = (e_1, \ldots, e_\ell)$ by $\lfloor \ell/3\rfloor$
cycles $(e_{3j-2}, e_{3j-1}, e_{3j})$ for $1 \leq j \leq \lfloor \ell/3\rfloor$.
If $\ell$ is not divisible by three, then we add one or two edges
$e_{\ell+1}, e_{\ell+2}$ to form a cycle of length three with the remaining
edge(s). Again, every $\alpha$-decomposition of the new cycle cover yields an
$\alpha$-decomposition of the original cycle cover.

In the remainder of this section, we assume that all directed cycle covers
consist solely of cycles of length two and all undirected cycle covers consist
solely of cycles of length three. Both theorems are proved using the
probabilistic method.

\subsubsection{Undirected Cycle Covers}
\label{sssec:existucc}

For the proof of Theorem~\ref{thm:alphau} below, we use Hoeffding's
inequality~\cite[Theorem 2]{Hoeffding:SumsBounded:1963}, which we state here in
a slightly modified version.

\begin{lemma}[Hoeffding's inequality]
\label{lem:hoeffding}
Let $X_1, \ldots, X_n$ be independent random variables, where $X_j$ assumes
values in $[a_j, b_j]$. Let $X = \sum_{j = 1}^n X_j$. Then
\[
  \probab\bigl(X < \expected(X) - t\bigr) \leq
  \exp\left(-\frac{2 t^2}{\sum_{j = 1}^n (b_j-a_j)^2}\right).
\]
\end{lemma}

\begin{theorem}
\label{thm:alphau}
For all $k \geq 2$, we have $\alpha_k^u \geq \frac 1{k}$.
\end{theorem}

\begin{proof}
Let $C$ be any cycle cover and $w_1, \ldots, w_k$ be $k$ objective functions.
First, we scale the edge weight such that $w_i(C) = k$ for all $i$. Thus,
$w_i(e) \leq 1$ for all edges $e$ of $C$ since the weight of any edge is at most
a $1/k$ fraction of the total weight. Second, we can assume that $C$ consists
solely of cycles of length three.

Let $c_1, \ldots, c_m$ be the cycles of $C$ and let $e_j^1, e_j^2, e_j^3$ be the
three edges of $c_j$. We perform the following random experiment: We remove one
edge of every cycle independently and uniformly at random to obtain a
decomposition $P$. Fix any $i \in [k]$. Let $X_j$ be the weight with respect to
$w_i$ of the path in $P$ that consists of the two edges of $c_j$. Then
$\expected(X_j) = 2w_i(c_j)/3$. Let $X = \sum_{j = 1}^m X_j$. Then
$\expected(w_i(X)) = 2w_i(C)/3 = 2k/3$.

Every $X_j$ assumes values between $a_j = \min\{w_i(e_j^1) + w_i(e_j^2),
w_i(e_j^1) + w_i(e_j^3), w_i(e_j^2) + w_i(e_j^3)\}$ and
$b_j = \max\{w_i(e_j^1) + w_i(e_j^2),
w_i(e_j^1) + w_i(e_j^3), w_i(e_j^2) + w_i(e_j^3)\}$. Since the weight of
each edge is at most $1$, we have $b_j - a_j \leq 1$.
Since the sum of all edge weights is $k$, we have
\[
  k \geq \sum_{j = 1}^m b_j \geq \sum_{j = 1}^m b_j - a_j \geq 
  \sum_{j = 1}^m (b_j - a_j)^2.
\]
Let us estimate the probability of the event that $X <1$, which corresponds to
$w_i(P) < 1$. If $\probab(X < 1) < 1/k$, then, by a union bound, we have
$\probab(\exists i: w_i(P) < 1) < 1$. Thus, $\probab(\forall i: w_i(P)\geq 1)
> 0$, which implies the existence of a $1/k$-decomposition. By Hoeffding's
inequality,
\[
       \probab(X < 1)
  =    \probab\left(X < \frac{2k}3 - \left(\frac{2k}3 - 1\right)\right)
  \leq \exp\left(-\frac{2 (\frac{2k}3 - 1)^2}{k}\right) =: p_k.
\]
We have $p_4 \approx 0.2494$, $p_5 \approx 0.11$, and $p_6 \approx 0.05$. Thus,
for $k = 4,5,6$, and also for all larger values of $k$, we have $p_k < 1/k$,
which implies the existence of a $1/k$-decomposition for $k \geq 4$. The cases
$k = 2$ and $k = 3$ remain to be considered since $p_3 \approx 0.51 > 1/3$ and
$p_2 \approx 0.89 > 1/2$. The bound for $\alpha_2^u$ follows from
Lemma~\ref{lem:alpha2u} below, which does not require
$w_i(e) \leq \alpha_2^u \cdot w_i(C)$.

Let us show $\alpha_3^u \geq 1/3$. This is done in a constructive way. First, we
choose from every cycle $c_j$ the edge $e_j^\ell$ that maximizes $w_3$ and put
it into $P'$. The set $P'$ will become a subset of $P$. Then $w_3(P') \geq 1$.
But we can also have some weight with respect to $w_1$ or $w_2$. Let
$\delta_1 = w_1(P')$ and $\delta_2 = w_2(P')$. If $\delta_i \geq 1$, then $w_i$
does not need any further attention.

Let $C' = C \setminus  P'$. We have $w_i(C') = 3-\delta_i$ for $i = 1,2$, and
$C'$ consists solely of paths of length two. Of every such path, we can choose
at most one edge for inclusion  in $P$. (Choosing both would create a cycle.)
Let $e_j^1, e_j^2$ be the two edges of $c_j$ with $w_2(e_j^2) \geq w_2(e_j^1)$.
Now we proceed by considering only $w_2$. Let $Q, Q'$ be initially empty sets.
For all $j = 1, \ldots, m$, if $w_2(Q) \geq w_2(Q')$, then we put (the heavier
edge) $e_j^2$ into $Q'$ and $e_j^1$ into $Q$. If $w_2(Q) \leq w_2(Q')$, then we
put $e_j^2$ into $Q$ and $e_j^1$ into $Q'$.

Both $P' \cup Q$ and $P' \cup Q'$ are decompositions of $C$. We claim that at
least one has a weight of at least $1$ with respect to all three objectives.
Since $w_3(P') \geq 1$, this holds for both with respect to $w_3$. Furthermore,
$|w_2(Q) - w_2(Q')| \leq 1$ since $w_2(e) \leq 1$ for all edges. We have
$w_2(Q) + w_2(Q') = 3-\delta_2$. Thus, $\min\{w_2(Q), w_2(Q')\} \geq
\frac{3-\delta_2}2 - \frac 12 \geq 1- \frac{\delta_2}2$. This implies
$w_2(P' \cup Q) \geq 1$ and $w_2(P' \cup Q') \geq 1$. Hence, with respect to
$w_2$ and $w_3$, both $P' \cup Q$ and $P' \cup Q$ will do. The first objective
$w_1$ remains to be considered. We have
$\max\{w_1(Q), w_1(Q')\} \geq \frac{3-\delta_1}2$. Choosing either
$P = P' \cup Q$ or $P = P' \cup Q'$ results in
$w_1(P) \geq \delta_1 + \frac{3-\delta_1}2 \geq 1$.
\end{proof}

For undirected graphs and $k = 2$, we do not need the assumption that the weight
of each edge is at most $\alpha_2^u$ times the weight of the cycle cover.
Lemma~\ref{lem:alpha2u} below immediately yields a $(1/2 - \eps)$ approximation
for bi-criteria \maxstsp: First, we compute a Pareto curve of cycle covers.
Second, we decompose each cycle cover to obtain a collection of paths, which we
then connect to form Hamiltonian cycles. The following lemma can also be
generalized to arbitrary $k$ (Lemma~\ref{lem:kmo}).

\begin{lemma}
\label{lem:alpha2u}
For every undirected cycle cover $C$ with edge weights $w=(w_1, w_2)$, there
exists a collection $P \subseteq C$ of paths with $w(P) \geq w(C)/2$.
\end{lemma}

\begin{proof}
Let $c$ be a cycle of $C$ consisting of edges $e_1, e_2, e_3$. Since we have
three edges, there exists one edge $e_j$ that is neither the maximum-weight edge
with respect to $w_1$ nor the maximum-weight edge with respect to $w_2$. We
remove this edge. Thus, we have removed at most half of $c$'s weight with
respect to either objective. Consequently, we have kept at least half of $c$'s
weight, which proves $\alpha_2^u \geq 1/2$.
\end{proof}

\subsubsection{Directed Cycle Covers}
\label{sssec:existdcc}

For directed cycle covers, our aim is again to show that the probability of
having not enough weight in one component is less than $1/k$. Hoeffding's
inequality works only for $k \geq 7$. We use a different approach, which
immediately gives us the desired result for $k \geq 6$, and which can be tweaked
to work also for small $k$.

\begin{theorem}
\label{thm:alphad}
For all $k \geq 2$, we have $\alpha_k^d \geq \frac 1{k+1}$.
\end{theorem}

\begin{proof}
As argued above, we can restrict ourselves to cycle covers consisting solely of
cycles of length two. We scale the edge weights to achieve $w_i(C) = k+1$ for
all $i \in [k]$. This implies $w_i(e) \leq 1$ for all edges $e \in C$.

Of every cycle, we randomly choose one of the two edges and put it into $P$. Fix
any $i \in [k]$. Our aim is to show that $\probab(w_i(P) < 1) < 1/k$, which
would prove the existence of an $\alpha_k^d$-decomposition. Let
$c_1, \ldots, c_m$ be the cycles of $C$ with $c_j=(e_j,f_j)$. Let
$w_i(e_j) = a_j$ and $w_i(f_j) = b_j$. We assume $a_j \leq b_j$ for all
$j \in [m]$. Let $\delta = \sum_{j = 1}^m a_j$. Then, no matter which edges we
choose, we obtain a weight of at least $\delta$. Hence, if $\delta \geq 1$, we
are done. Otherwise, we have $\delta < 1$ and replace $b_j$ by $b_j - a_j$ and
$a_j$ by $0$. Then we only need additional weight $1-\delta$, and our new goal
is to prove $\probab\bigl(w_i(P) < 1- \delta\bigr) < 1/k$.

This boils down to the following random experiment: We have numbers
$b_1, \ldots, b_m \in [0,1]$ with $\sum_{j=1}^m b_j = k+1 - 2 \delta$. Then we
choose a set $I \subseteq [m]$ uniformly at random. For such an $I$, we define
(by abusing notation) $w(I) = \sum_{j \in I} b_j$. We have to show
$\probab\bigl(w(I) < 1 -\delta\bigr) < 1/k$.

To this aim, let $C_1, \ldots, C_z \subseteq [m]$ with
$z = \bigl\lceil \frac{k+1}2 \bigr\rceil$ be pairwise disjoint sets with
$w(C_\ell) \in [1-\delta, 2-\delta)$. Such sets exist: We select arbitrary
elements for $C_1$ until $w(C_1) \in [1-\delta,2-\delta)$. This can always be
done since $b_j \leq 1$ for all $j$. Then we continue with $C_2$, $C_3$, and so
on. If we have already $z-1$ such sets, then
\[
       w(C_1 \cup \ldots \cup C_{z-1})
  \leq (2-\delta) \cdot (z-1) \leq (2-\delta) \cdot \frac k2 \leq k  - \delta
\]
since $k \geq 2$. Thus, at least weight $k+1-2 \delta - (k-\delta) = 1-\delta$
is left, which suffices for $C_z$.

The sets $C_1, \ldots, C_z$ do not necessarily form a partition of $[m]$. Let
$C' = [m] \setminus (C_1 \cup \ldots \cup C_z)$. We will have to consider $C'$
once in the end of the proof.

Now consider any $I, J \subseteq [m]$. We say that $I \sim J$ if
\[
  I = J \triangle C_{\ell_1} \triangle C_{\ell_2} \triangle \ldots \triangle
      C_{\ell_y}
\]
for some $C_{\ell_1}, \ldots, C_{\ell_y}$. Here, $\triangle$ denotes the
symmetric difference of sets. The relation $\sim$ is an equivalence relation
that partitions all subsets of $[m]$ into $2^{m-z}$ equivalence classes, each of
cardinality $2^z$. Let $[I] = \{J \subseteq [m] \mid J \sim I\}$.

\begin{lemma}
\label{lem:atmosttwo}
For every $I \subseteq [m]$, there are at most two sets $J \in [I]$ with
$w(J) < 1-\delta$.
\end{lemma}

\begin{proof}
Without loss of generality assume that $w(I) = \min_{J \in [I]} w(J)$. If
$w(I) \geq 1-\delta$, then there is nothing to show. Otherwise, consider any
$J = I \triangle C_{\ell_1} \triangle \ldots \triangle C_{\ell_y} \in [I]$ with
$y \geq 2$:
\[
       w(J)
  \geq \sum_{p=1}^y w(C_{\ell_p} \setminus I)
  \geq   \underbrace{\sum_{p=1}^y w(C_{\ell_p})}_{\geq y \cdot (1-\delta)
                                                  \geq 2-2\delta}
       - \underbrace{\sum_{p=1}^y w(C_{\ell_p} \cap I)}_{\leq w(I) < 1-\delta}
  \geq 1 - \delta.
\]
We conclude that the only possibility for other sets $J \in [I]$ with
$w(J) < 1-\delta$ is $J = I \triangle C_\ell$ for some $\ell$. We prove that
there is at most one such set by contradiction. So assume that there are
$J_1 = I \triangle C_1$ and  $J_2 = I \triangle C_2$ with
$w(J_1), w(J_2) < 1-\delta$. Then
$w(J_1) \geq w(C_1) - w(C_1 \cap I) + w(C_2 \cap I)$ and
$w(J_2) \geq w(C_2) - w(C_2 \cap I) + w(C_1 \cap I)$. Thus,
\[
  2 - 2\delta > w(J_1) + w(J_2) \geq w(C_1)  + w(C_2) \geq 2 - 2 \delta,
\]
a contradiction.
\end{proof}

A consequence of Lemma~\ref{lem:atmosttwo} is
$\probab\bigl(w(I) < 1-\delta\bigr) < 2^{-z+1} =
2^{-\lceil \frac{k+1}2 \rceil +1}$. This is less than $1/k$ for $k \geq 6$. The
cases $k \in \{2,3,4,5\}$ need special treatment.

Let us treat $k \in \{2,4\}$ first. Here
$2^{-\lceil \frac{k+1}2 \rceil +1} = 1/k$, which is almost good enough. To prove
$\probab\bigl(w(I) < 1-\delta\bigr) < 1/k$, we only have to find a set $I$ such
that at most one set $J \in [I]$ has $w(J) < 1 - \delta$. We claim that
$\emptyset$ is such a set: Of course $w(\emptyset) = 0 < 1-\delta$. But for any
other $J \in [\emptyset]$, we have
\[
  J = \emptyset \triangle C_{\ell_1} \triangle \ldots \triangle C_{\ell_y}
    = C_{\ell_1} \cup \ldots \cup C_{\ell_y}
\]
for some $C_{\ell_1}, \ldots, C_{\ell_y}$. The latter equality holds since the
sets $C_1, \ldots, C_z$ are disjoint. Thus
$w(J) \geq y \cdot (1-\delta) \geq 1-\delta$.

To finish the proof, we consider the case $k \in \{3,5\}$. For this purpose, we
consider $I$ and $\overline I = [m] \setminus I$ simultaneously. The classes
$[I]$ and $[\overline I]$ are disjoint: $[I] = [\overline I]$ would imply
$C' = \emptyset$. Then $\sum_{\ell=1}^z w(C_\ell) = k+1 - 2 \delta$. Since $k$
is odd, we have $z = \frac{k+1}2$. Thus, since $k+1 \geq 4$, there must exist an
$\ell$ with
\[
  w(C_\ell) \geq \frac{k+1-2\delta}{\frac{k+1}2}
            \geq \frac{(k+1) \cdot (2-\delta)}{k+1}
            =    2-\delta,
\]
which contradicts $w(C_\ell) < 2-\delta$.

We show that the number of sets $J \in [I] \cup [\overline I]$ with
$w(J)<1-\delta$ is at most two. This would prove the result for $k \in \{3,5\}$
since this would improve the bound to $\probab(w(I) < 1-\delta) < 2^{-(z+1)+1} =
2^{-\lceil \frac{k+1}2 \rceil} < 1/k$.

If we had more than two sets $J \in [I] \cup [\overline I]$ with
$w(J) < 1-\delta$, we can assume that we have two such sets in $[I]$. (We cannot
have more than two such $J$ due to Lemma~\ref{lem:atmosttwo}.) We assume that
these two sets are $I$ and $I' = I \triangle C_1$. Now consider any $J \in [I]$.
Since $k$ is odd and $k+1$ is even, we have
\begin{align*}
  w(J) & \leq   \underbrace{\sum_{\ell = 2}^z w(C_\ell)}_{\leq (2-\delta)
                                                               \cdot (z-1)}
              + \underbrace{\max\{w(I), w(I')\}}_{<1-\delta}
          <   (2-\delta) \cdot \left\lceil \frac{k+1}2\right\rceil -1 \\
       &  =   (2-\delta) \cdot \left(\frac{k+1}2\right) -1
          =   k - \frac{k+1}2 \cdot \delta \leq k-\delta.
\end{align*}
Thus, all sets $J \in [I]$ have a weight of less than $k-\delta$. This implies
$w(\overline J) = k+1-2 \delta - w(J) > 1-\delta$  for all
$\overline J \in [\overline I]$. Thus, if $[I]$ contains two sets whose weight
is less than $1-\delta$, then $[\overline I]$ contains no such set. 
\end{proof}

\subsubsection{Improvements and Generalizations}
\label{sssec:improve}

To conclude this section, let us discuss some improvements of the results
of this section. First, as a generalization of Lemma~\ref{lem:alpha2u}, cycle
covers without cycles of length at most $k$ can be $1/2$-decomposed. This,
however, does not immediately yield an approximation algorithm since
finding maximum-weight cycle covers where each cycle must have a length of
at least $k$ is NP- and APX-hard for $k \geq 3$ in directed graphs and for
$k \geq 5$~\cite{Manthey:RestrictedCC:2008}.

\begin{lemma}
\label{lem:kmo}
Let $k \geq 1$, and let $C$ be an arbitrary cycle cover such that the length of
every cycle is at least $k+1$. Then there exists a collection $P \subseteq C$ of
paths with $w(P) \geq w(C)/2$.
\end{lemma}

\begin{proof}
The proof is similar to the proof of Lemma~\ref{lem:alpha2u}. Let $c$ be any
cycle of $C$. For each $i \in [k]$, we choose one edge of $c$ that maximizes
$w_i$ for inclusion in $P$. Since $c$ has at least $k+1$ edges, this leaves us
(at least) one edge for removal.
\end{proof}

\begin{figure}[t]
\centering
\subfigure[$\alpha_2^d \leq 1/3$.]{\label{sfig:d2}%
          \includegraphics{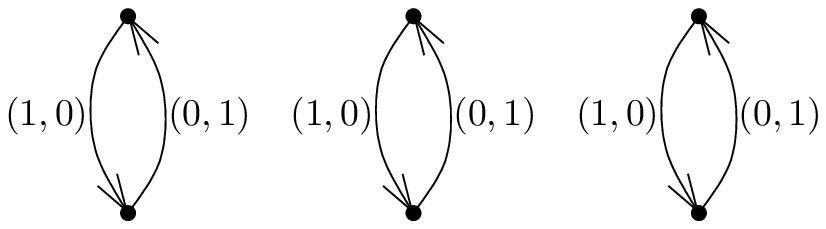}} \qquad \qquad
\subfigure[$\alpha_2^u \leq 1/2$.]{\label{sfig:u2}%
          \includegraphics{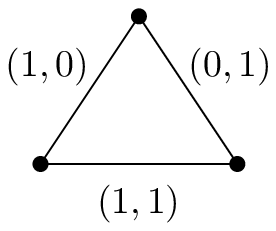}}
\caption{Examples that limit the possibility of decomposition.}
\label{fig:lowerdecomp}
\end{figure}

Figure~\ref{fig:lowerdecomp} shows that Theorems~\ref{thm:alphau}
and~\ref{thm:alphad}, respectively, are tight for $k = 2$. Due to these
limitations for $k = 2$, proving larger values for $\alpha_k^u$ or $\alpha_k^d$
does not immediately yield better approximation ratios (see
Section~\ref{sec:remarks}). However, for larger values of $k$, Hoeffding's
inequality yields the existence of $\Omega(1/\log k)$-decompositions. Together
with a different technique for heavy-weight cycle covers, this might lead to
improved approximation algorithms for larger values of $k$.

\begin{lemma}
\label{lem:suffk}
We have $\alpha_k^u, \alpha_k^d \in \Omega(1/\log k)$.
\end{lemma}

\begin{proof}
Let $A = c \ln k + d$ for some sufficiently large constants $c$ and $d$. Since
$\alpha_k^u \geq \alpha_k^d$, we can restrict ourselves to directed graphs.
Using the notation of Theorem~\ref{thm:alphau}, we have to show that
$\probab(X < 1) < 1/k$, where $X = \sum_{j = 1}^m X_j$ and
$X_j$ assumes values in the interval $[a_j, b_j]$, $b_j \leq a_j+1$,
$\sum_{j  = 1}^m (b_j + a_j)^2 \leq A$, and $\expected(X) = A/2$. We use
Hoeffding's inequality and plug in $t = A/2 -1$:
\[
  \probab(X < 1) \leq \exp\left(- \frac{2 (\frac A2 - 1)^2}{A}\right) 
                   =  \exp\left(- \frac A2 + 2 - \frac 2A \right)
                   < \frac 1k.
\]
\end{proof}

\subsection{Finding Decompositions}
\label{ssec:finddecomp}

While we know that decompositions exist due to the previous section, we have to
find them efficiently in order to use them in our approximation algorithm. We
present a deterministic algorithm and a faster randomized algorithm for finding
decompositions.

\subsubsection{Deterministic Algorithm}

\lw\ (Algorithm~\ref{alg:decomp}) is a deterministic algorithm for finding a
decomposition. The idea behind this algorithm is as follows: First, we scale the
weights such that $w(C) = 1/\alpha$. Then $w(e) \leq 1$ for all edges $e \in C$.
Second, we normalize all cycle covers such that they consist solely of cycles of
length two (in case of directed graphs) or three (in case of undirected graphs).
Third, we combine very light cycles as long as possible. More precisely,
if there are two cycles $c$ and $c'$ such that $w'(c)\leq 1/2$ and
$w'(c') \leq 1/2$, we combine them to one cycle $\tilde c$ with
$w'(\tilde c) \leq 1$. The requirements for an $\alpha$-decomposition to exist
are still fulfilled. Furthermore, any $\alpha$-decomposition of $C'$ immediately
yields an $\alpha$-decomposition of $C$.

\begin{algorithm}[t]
\begin{algorithmic}[1]
\item[] $P \leftarrow \lightweight (C, w, k, \alpha)$
\Input cycle cover $C$, edge weights $w$, $k \geq 2$,
       $w(e) \leq \alpha \cdot w(C)$ for all $e \in C$
\Output a collection $P$ of paths
\State \label{line:wprime}obtain $w'$ from $w$ by scaling each component such
       that $w'_i(C) = 1/\alpha$ for all $i$
\State normalize $C$ to $C'$ as described in the text such that $C'$ consists
       solely of cycles of length three (undirected) or two (directed)
\While{there are cycles $c$ and $c'$ in $C'$ with $w'(c) \leq 1/2$
       and $w'(c') \leq 1/2$}
   \State combine $c$ and $c'$ to $\tilde c$ with
          $w'(\tilde c) = w'(c) + w'(c')$
   \State replace $c$ and $c'$ by $\tilde c$ in $C'$
\EndWhile \label{line:endnormal}
\State try all possible combinations of decompositions
\State choose one $P'$ that maximizes $\min_{i \in [k]} w'_i(P)$
\State translate $P' \subseteq C'$ back to obtain a decomposition
       $P \subseteq C$
\State return $P$
\end{algorithmic}
\caption{A deterministic algorithm for finding a decomposition.}
\label{alg:decomp}
\end{algorithm}

The proof of the following lemma follows immediately from the existence
of decompositions (Theorems~\ref{thm:alphau} and~\ref{thm:alphad}).

\begin{lemma}
\label{lem:finddet}
Let $k \geq 2$. Let $C$ be an undirected cycle cover and $w_1, \ldots, w_k$ be
edge weights such that $w(e) \leq \alpha_k^u \cdot w(C)$. Then
$\lightweight(C, w, k, \alpha_k^u)$ returns a collection $P$ of paths with
$w(P) \geq \alpha_k^u \cdot w(C)$.

Let $C$ be a directed cycle cover and $w_1, \ldots, w_k$ be edge weights such
that $w(e) \leq \alpha_k^d \cdot w(C)$. Then $\lightweight(C, w, k, \alpha_k^d)$
returns a collection $P$ of paths with $w(P) \geq \alpha_k^d \cdot w(C)$.
\end{lemma}

Let us also estimate the running-time of \lw. The normalization in
lines~\ref{line:wprime} to~\ref{line:endnormal} can be implemented to run
in linear time. Due to the normalization, the weight of every cycle is at
least $1/2$ with respect to at least one $w_i'$. Thus, we have at most
$2k/\alpha_k^u$ cycles in $C'$ in the undirected case and at most
$2k/\alpha_k^d$ cycles in $C'$ in the directed case. In either case, we have
$O(k^2)$ cycles. All of these cycles are of length two or of length three. Thus,
we find an optimal decomposition, which in particular is an $\alpha_k^u$ or
$\alpha_k^d$-decomposition, in time linear in the input size and exponential in
$k$.

\subsubsection{Randomized Algorithm}

By exploiting the probabilistic argument of the previous section, we can find a
decomposition much faster with a randomized algorithm. \rlw\
(Algorithm~\ref{alg:randecomp}) does this: We choose the edges to be deleted
uniformly at random for every cycle. The probability that we obtain a
decomposition as required is positive and bounded from below by a constant.
Furthermore, as the proofs of Theorems~\ref{thm:alphau} and~\ref{thm:alphad}
show, this probability tends to one as $k$ increases. For $k \geq 6$, it is at
least approximately $0.7$ for undirected cycle covers and at least $1/4$ for
directed cycle covers. For $k < 6$, we just use our deterministic algorithm,
which has linear running-time for constant $k$. The following lemma follows from
the considerations above.

\begin{algorithm}[t]
\begin{algorithmic}[1]
\item[] $P \leftarrow \randlightweight (C, w, k, \alpha)$
\Input cycle cover $C$, edge weights $w = (w_1, \ldots, w_k)$, $k \geq 2$,
       $w(e) \leq \alpha \cdot w(C)$ for all $e \in C$
\Output a collection $P$ of paths with $w(P) \geq \alpha \cdot w(C)$
\If{$k \geq 6$}
   \Repeat
      \State randomly choose one edge of every cycle of $C$
      \State remove the chosen edges to obtain $P$
   \Until{$w(P) \geq \alpha \cdot w(C)$}
\Else
   \State $P \leftarrow \lightweight (C, w, k, \alpha)$
\EndIf
\end{algorithmic}
\caption{A randomized algorithm for finding a decomposition.}
\label{alg:randecomp}
\end{algorithm}

\begin{lemma}
\label{lem:findrand}
Let $k \geq 2$. Let $C$ be an undirected cycle cover and $w_1, \ldots, w_k$ be
edge weights such that $w(e) \leq \alpha_k^u \cdot w(C)$. Then
$\randlightweight(C, w, k, \alpha_k^u)$ returns a collection $P$ of paths with
$w(P) \geq \alpha_k^u \cdot w(C)$.

Let $C$ be a directed cycle cover and $w_1, \ldots, w_k$ be edge weights such
that $w(e) \leq \alpha_k^d \cdot w(C)$. Then
$\randlightweight(C, w, k, \alpha_k^d)$ returns a collection $P$ of paths with
$w(P) \geq \alpha_k^d \cdot w(C)$.

The expected running-time of \rlw\ is $O(|C|)$.
\end{lemma}

\section{Approximation Algorithms}
\label{sec:alg}

Based on the idea sketched in Section~\ref{sec:idea}, we can now present our
approximation algorithms for multi-criteria \maxatsp\ and \maxstsp. However, in
particular for \maxstsp, some additional work has to be done if heavy edges are
present.

\subsection{Multi-Criteria \maxatsp}
\label{sec:algatsp}

We first present our algorithm for \maxatsp\ (Algorithm~\ref{alg:maxatsp}) since
it is a bit easier to analyze.

\begin{algorithm}[t]
\begin{algorithmic}[1]
\item[] $\ptsp \leftarrow \atspalg(G, w, k, \eps)$
\Input directed complete graph $G=(V,E)$, $k \geq 1$, edge
       weights $w: E \to \nat^k$, $\eps > 0$
\Output approximate Pareto curve \ptsp\ for $k$-criteria Max-TSP
\If{$k = 1$}
    \State compute a $2/3$ approximation $\ptsp$
\Else
   \State compute a $(1-\eps)$ approximate Pareto curve $\pc$ of cycle
          covers
   \State $\ptsp \leftarrow \emptyset$
   \ForAll{cycle covers $C \in \pc$}
      \If{$w(e) \leq \alpha_k^d \cdot w(C)$ for all edges $e \in C$}
            \label{line:decide}
         \State $P \leftarrow \lightweight (C, w, k)$
         \State add edges to $P$ to form a Hamiltonian cycle $H$
         \State add $H$ to $\ptsp$
      \Else
         \State let $e = (u,v) \in C$ be an edge with
                $w(e) \not\leq \alpha_k^d \cdot w(C)$
         \ForAll{$a, b, c, d \in V$ such that $P_{a,b,c,d}^e$ is legal}
            \For{$i \leftarrow 1 \text{\textbf{ to }} k$}
               \State obtain $G'$ from $G$ by contracting the paths of
                      $P_{a,b,c,d}^e$
               \State obtain $w'$ from $w$ by removing the $i$th objective
               \State $\ptsp' \leftarrow \atspalg(G', w', k-1, \eps)$
               \ForAll{$H' \in \ptsp'$}
                  \State form a Hamiltonian cycle from $H'$ plus
                         $P_{a,b,c,d}^e$; add it to $\ptsp$
                  \State form a Hamiltonian cycle from $H'$ plus $(u,v)$; add it
                         to $\ptsp$
               \EndFor
            \EndFor
         \EndFor
      \EndIf
   \EndFor
\EndIf
\end{algorithmic}
\caption{Approximation algorithm for $k$-criteria \maxatsp.}
\label{alg:maxatsp}
\end{algorithm}

First of all, we compute a $(1-\eps)$ approximate Pareto curve $\pc$ of cycle
covers. Then, for every cycle cover $C \in \pc$, we decide whether it is a
light-weight cycle cover or a heavy-weight cycle cover (line~\ref{line:decide}).
If $C$ has only light-weight edges, we decompose it to obtain a collection $P$
of paths. Then we add edges to $P$ to obtain a Hamiltonian cycle $H$, which we
then add to $\ptsp$.

\begin{figure}[t]
\centering
\subfigure[Two disjoint paths.]{\label{sfig:twopaths}%
           \includegraphics{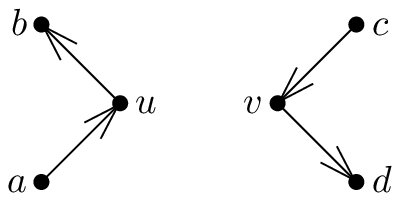}} \quad
\subfigure[One path with an intermediate vertex.]{\label{sfig:onelong}%
           ~\quad\qquad\includegraphics{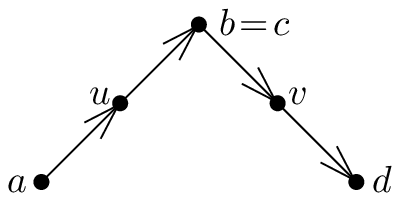}\qquad\quad~} \quad
\subfigure[One path including $e$.]{\label{sfig:oneshort}%
           \includegraphics{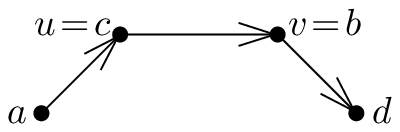}}
\caption{The three possibilities of $P_{a,b,c,d}^e$. Symmetrically to
         \subref{sfig:onelong}, we also have $a=d$. Symmetrically to
         \subref{sfig:oneshort}, we also have $v=a$ and $u=d$.}
\label{fig:pabcde}
\end{figure}

If $C$ contains a heavy-weight edge, then there exists an edge $e = (u,v)$ and
an $i$ with $w_i(e) > \alpha_k \cdot w_i(C)$. We pick one such edge. Then we
iterate over all possible vertices $a, b, c, d$ (including equalities and
including $u$ and $v$). We denote by $P_{a,b,c,d}^e$ the graph with vertices
$u$, $v$, $a$, $b$, $c$, $d$ and edges $(a,u)$, $(u,b)$, $(c,v)$, and $(v,d)$.
We call $P_{a,b,c,d}^e$ \emph{legal} if it can be extended to a Hamiltonian
cycle: $P_{a,b,c,d}^e$ is legal if and only if it consists of one or two
vertex-disjoint directed paths. Figure~\ref{fig:pabcde} shows the different
possibilities.

For every legal $P_{a,b,c,d}^e$, we contract the paths as follows: We remove
all outgoing edges of $a$ and $c$, all incoming edges of $b$ and $d$, and all
edges incident to $u$ or $v$. Then we identify $a$ and $b$ as well as $c$ and
$d$. If $P_{a,b,c,d}^e$ consists of a single path, then we remove all vertices
except the two endpoints of this path, and we identify these two endpoints.

In this way, we obtain a slightly smaller instance $G'$. Then, for every $i$, we
remove the $i$th objective to obtain $w'$, and recurse on $G'$ with only $k-1$
objectives $w'$. This yields approximate Pareto curves $\ptsp'$ of Hamiltonian
cycles of $G'$. Now consider any $H' \in \ptsp'$. We expand the contracted paths
to obtain $H$. Then we construct two tours: First, we just add $P_{a,b,c,d}^e$
to $H$, which yields a Hamiltonian cycle by construction. Second, we observe
that no edge in $H$ is incident to $u$ or $v$. We add the edge $(u,v)$ to $H$ as
well as some more edges such that we obtain a Hamiltonian cycle. We put the
Hamiltonian cycles thus constructed into $\ptsp$.

We have not yet discussed the success probability. Randomness is needed for
computing the approximate Pareto curves of cycle covers and the recursive calls
of \apa\ with $k-1$ objectives. Let $N$ be the size of the instance at hand, and
let $p_k(N, 1/\eps)$ is a polynomial that bounds the size of a $(1-\eps)$
approximate Pareto curve from above. Then we need at most
$N^4 \cdot p_k(N, 1/\eps)$ recursive calls of \apa. In total, the number of
calls of randomized algorithms is bounded by some polynomial $q_k(N, 1/\eps)$.
We amplify the success probabilities of these calls such that the probability is
at least $1-\frac 1{2 \cdot q_k(N, 1/\eps)}$. Thus, the probability that one
such call is not successful is at most
$q_k(N, 1/\eps) \cdot \frac 1{2 \cdot q_k(N, 1/\eps)} \leq 1/2$ by a union
bound. Hence, the success probability of the algorithm is at least $1/2$.

Instead of \lw, we can also use its randomized counterpart \rlw. We modify \rlw\
such that the running-time is guaranteed to be polynomial and that there is only
a small probability that \rlw\ errs. Furthermore, we have to make the error
probabilities of the cycle cover computation as well as the recursive calls of
\apa\ slightly smaller to maintain an overall success probability of at least
$1/2$.

Overall, the running-time of \apa\ is polynomial in the input size and $1/\eps$,
which can be seen by induction on $k$: We have a polynomial time approximation
algorithm for $k = 1$. For $k > 1$, the approximate Pareto curve of cycle covers
can be computed in polynomial time, yielding a polynomial number of cycle covers.
All further computations can also be implemented to run in polynomial time since
\apa\ for $k-1$ runs in polynomial time by induction hypothesis.

\begin{theorem}
\label{thm:aalg}
\apa\ is a randomized $\frac{1}{k+1} - \eps$ approximation for multi-criteria
\maxatsp. Its running-time is polynomial in the input size and $1/\eps$.
\end{theorem}

\begin{proof}
We have already discussed the error probabilities and the running-time. Thus, it
remains to consider the approximation ratio, and we can assume in the following,
that all randomized computations are successful. We prove the theorem by
induction on $k$. For $k = 1$, this follows since mono-criterion \maxatsp\ can
be approximated with a factor $2/3 > 1/2$.

Now assume that the theorem holds for $k-1$. We have to prove that, for every
Hamiltonian cycle $\hat H$, there exists a Hamiltonian cycle $H \in \ptsp$ with
$w(H) \geq \bigl(\frac{1}{k+1} - \eps\bigr) \cdot w(\hat H)$. Since every
Hamiltonian cycle is in particular a cycle cover, there exists a $C \in \pc$
with $w(C) \geq (1-\eps) \cdot w(\hat H)$. Now we distinguish two cases.

The first case is that $C$ consists solely of light-weight edges, i.~e.,
$w(e) \leq \frac{1}{k+1} \cdot w(C)$, then \lw\ returns a collection $P$ of
paths with $w(P) \geq \frac{1}{k+1} \cdot w(C) \geq
\bigl(\frac{1}{k+1}- \eps\bigr) \cdot w(\hat H)$, which yields a Hamiltonian
cycle $H$ with
$w(H) \geq w(P) \geq \bigl(\frac{1}{k+1}- \eps\bigr) \cdot w(\hat H)$ as
claimed.

The second case is that $C$ contains at least one heavy-weight edge $e = (u,v)$.
Let $(a,u)$, $(u,b)$, $(c, v)$, and $(v, d)$ be the edges in $\hat H$ that are
incident to $u$ or $v$. (We may have some equalities among the vertices as shown
in Figure~\ref{fig:pabcde}.) Note that $\hat H$ does not necessarily contain the
edge $e$. We consider the corresponding $P_{a,b,c,d}^e$ and divide the second
case into two subcases.

The first subcase is that there exists a $j \in [k]$ with
$w_j(P_{a,b,c,d}^e) \geq \frac{1}{k+1} \cdot w_j(\hat H)$, i.~e., at least a
$\frac{1}{k+1}$ fraction of the $j$th objective is concentrated in
$P_{a,b,c,d}^e$. (We can have $j=i$, but this is not necessarily the case.) Let
$J \subseteq [k]$ be the set of such $j$.

We fix one $j \in J$ arbitrarily and consider the graph $G'$ obtained by
removing the $j$th objective and contracting the paths $(a,u,b)$ and $(c,v,d)$.
A fraction of $1-\frac 1{k+1} = \frac k{k+1}$ of the weight of $\hat H$ is left
in $G'$ with respect to all objectives but those in $J$. Thus, $G'$ contains a
Hamiltonian cycle $\hat H'$ with
$w_{\ell}(\hat H') \geq \frac{k}{k+1} \cdot w_{\ell}(\hat H)$ for all
$\ell \in [k] \setminus J$. Since $(k-1)$-criteria \maxatsp\ can be approximated
with a factor of $\frac 1k - \eps$ by assumption, $\ptsp'$ contains a
Hamiltonian cycle $H'$ with $w_\ell(H') \geq
(\frac 1k - \eps) \cdot \frac{k}{k+1} \cdot w_\ell(\hat H) \geq
\bigl(\frac{1}{k+1} -\eps\bigr) \cdot w_\ell(\hat H)$ for all
$\ell \in [k] \setminus J$. Together with $P_{a,b,c,d}^e$, which contributes
enough weight to the objectives in $J$, we obtain a Hamiltonian cycle $H$ with
$w(H) \geq \bigl(\frac{1}{k+1} -\eps\bigr) \cdot w(\hat H)$, which is as
claimed.

The second subcase is that $w_j(P_{a,b,c,d}^e) \leq \frac{1}{k+1} \cdot w_j(H)$
for all $j \in [k]$. Thus, at least a fraction of $\frac k{k+1}$ of the weight
of $\hat H$ is outside of $P_{a,b,c,d}^e$. We consider the case with the $i$th
objective removed. Then, with the same argument as in the first subcase, we
obtain a Hamiltonian cycle $H'$ of $G'$ with $w_\ell(H') \geq
\bigl(\frac{1}{k+1} -\eps\bigr) \cdot w_\ell(\hat H)$ for all
$\ell \in [k] \setminus \{i\}$. To obtain a Hamiltonian cycle of $G$, we take
the edge $e = (u,v)$ and connect its endpoints appropriately. (For instance, if
$a,b,c,d$ are distinct, then we add the path $(a,u,v,d)$ and the edge $(c,b)$.)
This yields enough weight for the $i$th objective in order to obtain a
Hamiltonian cycle $H$ with
$w(H) \geq \bigl(\frac{1}{k+1}- \eps\bigr) \cdot w(\hat H)$ since $w_i(e) \geq
\frac{1}{k+1} \cdot w(C) \geq \bigl(\frac{1}{k+1} - \eps\bigr) \cdot w(\hat H)$.
\end{proof}

\subsection{Multi-Criteria \maxstsp}
\label{sec:algstsp}

\apa\ works of course also for undirected graphs, for which it achieves an
approximation ratio of $\frac{1}{k+1} - \eps$. But we can do better for
undirected graphs.

Our algorithm \spa\ for undirected graphs (Algorithm~\ref{alg:maxstsp}) starts
by computing an approximate Pareto curve of cycle covers just as \apa\ did. Then
we consider each cycle cover $C$ separately. If $C$ consists solely of
light-weight edges, then we can decompose $C$ using \lw. If $C$ contains one or
more heavy-weight edges, then some more work has to be done than in the case of
directed graphs. The reason is that we cannot simply contract paths -- this
would make the new graph $G'$ (and the edge weights $w'$) asymmetric.

\begin{algorithm}[t]
\begin{algorithmic}[1]
\item[] $\ptsp \leftarrow \stspalg(G, w, k, \eps)$
\Input undirected complete graph $G=(V,E)$, $k \geq 2$, edge
       weights $w: E \to \nat^k$, $\eps > 0$
\Output approximate Pareto curve \ptsp\ for $k$-criteria Max-TSP
\State compute a $(1-\eps)$ approximate Pareto curve $\pc$ of cycle
       covers
\State $\ptsp \leftarrow \emptyset$
\If{$k = 2$}
    \ForAll{$C \in \pc$}
       \State $P \leftarrow \lw (C, w, k)$
       \State add edges to $P$ to form a Hamiltonian cycle $H$
       \State add $H$ to $\ptsp$
    \EndFor
\Else
   \ForAll{cycle covers $C \in \pc$}
      \If{$w(e) \leq w(C)/k$ for edges $e \in C$}
         \State $P \leftarrow \lightweight (C, w, k)$
         \State add edges to $P$ to form a Hamiltonian cycle $H$
         \State add $H$ to $\ptsp$
      \Else
         \State let $i \in [k]$ and $e = \{u,v\} \in C$ with
                $w_i(e) > w_i(C)/k$
         \ForAll{$\ell \in \{0, \ldots, 4k\}$, distinct
                 $x_1, \ldots, x_\ell \in V \setminus \{u,v\}$, and $k \in [k]$}
            \State $U \leftarrow \{x_1, \ldots, x_\ell, u, v\}$
            \State obtain $w'$ from $w$ by removing the $j$th objective
            \State set $w'(f) = 0$ for all edges $f$ incident to $U$
            \State $\ptsp^{U,j} \leftarrow \stspalg(G, w', k-1, \eps)$
            \ForAll{$H \in \ptsp^{U,j}$}
               \State remove all edges $f$ from $H$ with $f \subseteq U$
                      to obtain $H'$
               \ForAll{$H_U$ such that $H' \cup H_U$ is a Hamiltonian cycle}
                  \State add $H' \cup H_U$ to $\ptsp$
               \EndFor
            \EndFor
         \EndFor
      \EndIf
   \EndFor
\EndIf
\end{algorithmic}
\caption{Approximation algorithm for $k$-criteria \maxstsp.}
\label{alg:maxstsp}
\end{algorithm}

So assume that a cycle cover $C \in \pc$ contains a heavy-weight edge
$e = \{u,v\}$. Let $i \in [k]$ be such that $w_i(e) \geq w_i(C)/k$. In a first
attempt, we remove the $i$th objective to obtain $w'$. Then we set $w'(f) = 0$
for all edges $f$ incident to $u$ or $v$. We recurse with $k-1$ objectives on
$G$ with edge weights $w'$. This yields a tour $H'$ on $G$. Now we remove all
edges incident to $u$ or $v$ of $H'$ and add new edges including $e$. In this
way, we get enough weight with respect to objective $i$. Unfortunately, there is
a problem if there is an objective $j$ and an edge $f$ incident to $u$ or $v$
such that $f$ contains almost all weight with respect to $w_j$: We cannot
guarantee that this edge $f$ is included in $H$ without further modifying
$H'$. To cope with this problem, we do the following: In addition to $u$ and
$v$, we set the weight of all edges incident to the other vertex of $f$ to $0$.
Then we recurse. Unfortunately, there may be another objective $j'$ that now
causes problems. To solve the whole problem, we iterate over all
$\ell = 0, \ldots, 4k$ and over all additional vertices
$x_1, \ldots, x_\ell \neq u,v$. Let $U = \{x_1, \ldots, x_\ell, u, v\}$. We
remove one objective $i \in [k]$ to obtain $w'$, set the weight of all edges
incident to $U$ to $0$, and recurse with $k-1$ objectives. Although the time
needed to do this is exponential in $k$, we maintain polynomial running-time
for fixed~$k$.

As in the case of directed graphs, we can make the success probability of every
randomized computation small enough to maintain a success probability of at
least $1/2$.

The base case is now $k=2$: In this case, every cycle cover possesses a $1/2$
decomposition, and we do not have to care about heavy-weight edges. Overall, we
obtain the following result.

\begin{theorem}
\label{thm:salg}
\spa\ is a randomized $\frac 1k - \eps$ approximation for multi-criteria
\maxstsp. Its running-time is polynomial in the input size and $1/\eps$.
\end{theorem}

\begin{proof}
We have already dealt with error probabilities and running-time. Thus, we can
assume that all randomized computations are successful in the following. What
remains to be analyzed is the approximation ratio. As in the proof of
Theorem~\ref{thm:aalg}, the proof is by induction on $k$.

The base case is $k=2$. Let $\hat H$ be an arbitrary Hamiltonian cycle. Then
there is a $C \in \pc$ with $w(C) \geq (1-\eps) \cdot w(\hat H)$. From $C$, we
obtain a Hamiltonian cycle $H$ with
$w(H) \geq \frac 12 \cdot w(C) \geq (\frac 12 -\eps) \cdot w(\hat H)$ by
decomposition and Lemma~\ref{lem:alpha2u}.

Let us analyze \spa\ for $k \geq 3$ objectives. By the induction hypothesis, we
can assume that \spa\ is a $\frac{1}{k-1} - \eps$ approximation for
$(k-1)$-criteria \maxstsp. Let $\hat H$ be any Hamiltonian cycle. We have to
show that $\ptsp$ contains a Hamiltonian cycle $H$ with
$w(H) \geq \bigl(\frac 1k - \eps \bigr) \cdot w(\hat H)$.

There is a $C \in \ptsp$ with $w(C) \geq (1-\eps) \cdot w(\hat H)$. We have to
distinguish two cases. First, if $C$ consists solely of light-weight edges,
i.~e., $w(e) \leq w(C)/k$ for all $e$, then we obtain a Hamiltonian cycle $H$
from $C$ with $w(H) \geq w(C)/k \geq
\bigl(\frac 1k - \eps\bigr) \cdot w(\hat H)$.

Second, let $e \in C$ and $i \in [k]$ with $w_i(e) > w_i(C)/k$. We construct
sets $I \subseteq [k]$, $X \subseteq \hat H$, and $U \subseteq V$ in phases as
follows (we do not actually construct these sets, but only need them for the
analysis): Initially, $I = X = \emptyset$ and $U = \{u,v\}$. In every phase, we
consider the set $X'$ of all edges of $\hat H$ that have exactly one endpoint in
$U$. We always have $|X'| \leq 4$ by construction. Let
$I' = \{j \in [k] \mid j \notin I, w_j(X') \geq w_j(\hat H)/k\}$. If $I'$ is
empty, then we are done. Otherwise, add $I'$ to $I$, add $X'$ to $X$, and add
all new endpoints of vertices in $X'$ to $U$. We add at least one element to $I$
in every phase. Thus, $|X| \leq 4k$ and $|U| \leq 4k+2$ since $|I| \leq k$.

Let $w^\inside = w(X)$, and let
$w^\border = \sum_{f \in \hat H: |f \cap U| = 1} w(f)$ be the weight of edges of
$\hat H$ that have exactly one endpoint in $U$. Let
$w^\outside = w(\hat H) - w^\inside - w^\border$. By construction, we have
$w^\border_j < 1/k$ for all $j \notin I$. Otherwise, we would have added more
edges to $X$.

We distinguish two subcases. The first subcase is that $I = \emptyset$. Then
$w^\inside = 0$ and $w^\border < 1/k$. Consider the set $\ptsp^{\emptyset, i}$
and the edge weights $w'$ used to obtain it. We have
$w'_j(\hat H) = w^\outside_j > \bigl(\frac{k-1}k \bigr) \cdot w_j(\hat H)$ for
$j \neq i$. By the induction hypothesis, there is an
$H \in \ptsp^{\emptyset, i}$ with
\[
       w'_j(H)
  \geq       \left(\frac 1{k-1} - \eps\right) \cdot \left(\frac{k-1}k \right)
       \cdot w(\hat H)
  \geq \left(\frac 1{k} - \eps\right) \cdot w(\hat H)
\]
for $j \neq i$. We remove all edges incident to $u$ or $v$ to obtain $H'$. Since
the weight of all these edges has been set to $0$, we have $w'(H') = w'(H)$.
There exists a set $H_\emptyset$ such that $e \in H_\emptyset$ and
$H' \cup H_\emptyset$ is a Hamiltonian cycle. For this cycle, which is in
$\ptsp$, we have
\[
       w_i(H' \cup H_\emptyset) \geq w_i(e) \geq w_i(C)/k
  \geq \left(\frac 1k - \eps \right) \cdot w(\hat H)
\]
and, for $j \neq i$,
\[
       w_j(H' \cup H_\emptyset)
  \geq w'_j(H)
  \geq \left(\frac 1{k} - \eps\right) \cdot w(\hat H).
\]

The second subcase is that $I$ is not empty. Let $j \in I$, and let $U$. We
consider $\ptsp^{U,j}$. Let $w^\inside$, $w^\border$, and $w^\outside$ be as
defined above. By the induction hypothesis, the set $\ptsp^{U,j}$ contains a
Hamiltonian cycle cover $H$ with
$w'_\ell(H) \geq \bigl(\frac 1{k-1} - \eps \bigr) \cdot w^\outside_\ell$ for
$\ell \neq j$. We remove all edges incident to $U$ from $H$ to obtain $H'$ with
$w'(H') = w'(H)$. By construction $H' \cup X$ is a collection of paths. We add
edges to $X$ to obtain $H_U$ such that $H' \cup H_U$ is a Hamiltonian cycle. Let
us estimate the weight of $H' \cup H_U$. For all $\ell \in I$, we have
$w_\ell(H' \cup H_U) \geq w_\ell(H_U) \geq w_\ell(\hat H)/k$. For all
$\ell \notin I$, we have
\begin{align*}
          w_\ell(H' \cup H_U)
   & \geq w'_\ell(H') + w^\inside_\ell
     \geq      \left(\frac 1{k-1} - \eps \right)
          \cdot (w^\outside_\ell + w^\inside_\ell)
     \geq      \left(\frac 1{k-1} - \eps \right)
          \cdot (w_\ell(\hat H) - w^\border_\ell) \\
   & \geq       \left(\frac 1{k-1} - \eps \right)
          \cdot \frac{k-1}k \cdot w_\ell(\hat H)
     \geq \left(\frac 1{k} - \eps \right) \cdot w_\ell(\hat H),
\end{align*}
which completes the proof.
\end{proof}

\section{Remarks}
\label{sec:remarks}

The analysis of the approximation ratios of our algorithms is essentially
optimal: Our approach can at best lead to approximation ratios of
$\frac{1}{k+c}$ for some $c \in \integer$. The reason is as follows: Assume that
$(k-1)$-criteria \maxtsp\ can be approximated with a factor of $\tau_k$. If we
have a $k$-criteria instance, we have to set the threshold for heavy-weight
edges somewhere. Assume for the moment that this threshold $\alpha_k$ be
arbitrary. Then the ratio for $k$-criteria \maxtsp\ is
$\min\{\alpha_k, (1-\alpha_k) \cdot \tau_{k-1}\}$. Choosing
$\alpha_k = \frac{\tau_{k-1}}{\tau_{k-1}+1}$ maximizes this ratio. Thus, if
$\tau_{k-1} = 1/T$ for some $T$, then
$\tau_k \leq \frac{\tau_{k-1}}{\tau_{k-1}+1} = \frac{1}{T+1}$. We conclude that
the denominator of the approximation ratio increases by at least $1$ if we go
from $k-1$ to $k$.

For undirected graphs, we have obtained a ratio of roughly $1/k$, which is
optimal since $\alpha_2^u = 1/2$ implies $c \geq 0$. Similarly, for directed
graphs, we have a ratio of $\frac 1{k+1}$, which is also optimal since
$\alpha_2^d = 1/3$ implies $c \geq 1$.

Due to the existence of $\Omega(1/\log k)$-decompositions, we conjecture that
both $k$-criteria \maxstsp\ and $k$-criteria \maxatsp\ can in fact be
approximated with factors of $\Omega(1/\log k)$. This, however, requires a
different approach or at least a new technique for heavy-weight edges.


\end{document}